Energy loss by the arbitrary moving charged particle in classical electrodynamics

Lidsky V.V.

A B S T R A C T

The energy-impulse flow of electromagnetic field, produced by the arbitrary moving charged particle, through the sphere of large radius is calculated. The result formulae are used in problem of radiation produced by hyperbolically moving particle and the problem of the synchrotron radiation energy loss. The energy-impulse radiated by a moving charged particle is shown to be not an invariant 4-tensor, for this reason calculation methods based on change of reference frame are not correct.

Потери энергии заряженной частицей, движущейся по произвольной траектории, в классической электродинамике.

В. В. Лидский

А Н Н О Т А Ц И Я

Вычислен поток энергии электромагнитного поля, создаваемого произвольно движущейся заряженной частицей, через неподвижную удаленную сферу. Полученный результат применен к задаче об излучении частицы, движущейся по винтовой линии в магнитном поле, и к задаче об излучении при гиперболическом движении. Показано, что величина энергии-импульса, излучаемая частицей, не инвариантна, в связи с чем способы вычисления мощности излучения через переход к более удобной системе координат приводят к некорректным результатам.

**1. Введение.**

В 60-е шла дискуссия о методе вычисления потери энергии движущейся заряженной частицей. Итог был подведен Гинзбургом, Сазоновым и Сыроватским в работах [1]. Авторами была получена корректная формула для мощности магнитотормозного (или синхротронного) излучения. Там же можно найти обзор проблем, связанных с этим вопросом.

В нашей работе мы рассмотрим задачу о потере энергии заряженной частицей, движущейся по произвольной траектории.

Для вычисления мощности электромагнитного излучения частицы вычислим поток тензора энергии-импульса поля через поверхность неподвижной удаленной сферы $S$. Центр сферы поместим вблизи мировой линии частицы, а радиус R выберем достаточно большим. Будем предполагать, что поле частицы определяется запаздывающими потенциалами Лиенара-Вихерта. Соответствующие напряженности электромагнитного поля $F^{ik}$ представим в виде

(1.2) $$F^{ik} = \frac{e}{(\chi^j \cdot w_j)^3} \cdot (\chi^i \cdot w^k - w^i \cdot \chi^k) \cdot (1 - \chi^m a_m) + \frac{e}{(\chi^j \cdot w_j)^2} \cdot (\chi^i \cdot a^k - a^i \cdot \chi^k).$$

Здесь $w^i$, $a^i$ — 4-скорость и 4-ускорение частицы, а $\chi^i = x^i - z^i$ – светоподобный 4-вектор из «точки излучения» $z^i$ в «точку наблюдения» $x^i$.

Здесь и далее мы будем придерживаться обозначений, принятых в [2]. Латинские индексы пробегают значения $i=(0,1,2,3)$. Индекс тензора, обозначенный греческими буквами $\alpha,\beta,\gamma$ пробегает значения пространственных координат $(1,2,3)$. Опускание и поднятие индексов происходит с помощью свертки с метрическим тензором Минковского $g^{ik}=(+---)$. Этот же тензор используется и при поднятии индекса у трехмерных векторов, так что $x_\alpha = -x^\alpha$. Скорость света принята равной единице: $c=1$.

Для тензора энергии-импульса примем формулу Максвелла-Хевисайда [2]:

$$(1.3) \qquad T^{ik} = \frac{1}{4\pi}\left(-F^{il}F^{k}{}_{l} + \frac{1}{4}g^{ik}F_{lm}F^{lm}\right)$$

## 2. Преобразование поверхностного интеграла.

2.1. В этом параграфе мы получим формулу для полного потока тензора энергии-импульса поля Лиенара-Вихерта через удаленную сферу $S$, справедливую для частицы, движущейся по произвольной мировой линии. Систему отсчета наблюдателя, относительно которого сфера неподвижна будем называть лабораторной.

Нам будет удобно использовать вектор 4-скорости лабораторного наблюдателя

$$(2.1) \qquad u^i = \{1,0,0,0\}.$$

Момент времени, в который мы вычислим поток энергии через сферу $S$, примем равным радиусу сферы $R$, тогда «момент излучения» частицы окажется близким к моменту $t=0$.

Начало координат поместим в центр сферы, тогда 4-вектор центра примет вид:

$$(2.2) \qquad \gamma^j = \{0,0,0,0\}.$$

В этом случае 4-вектор $x^i$, соединяющий центр сферы $\gamma^j$ с некоторой точкой поверхности, окажется светоподобным:

$$(2.3) \qquad x^i \cdot x_i = 0$$

Мощность излучения частицы выражается через полный поток тензора энергии-импульса через сферу $S$:

$$(2.4) \qquad \frac{dP^i}{dt} = \oint_S T^{ik} \cdot n_k \cdot dS$$

где $n_k$ – 4-вектор внешней нормали к поверхности сферы:

$$(2.5) \qquad n_i = \frac{1}{R}\{0,-x_1,-x_2,-x_3\}$$

Временная компонента нормали обращается в ноль, так как сферу $S$ мы считаем неподвижной. Знак минус в (2.5) выбран для того, чтобы поток вектора, направленного вовне сферы был положительным (напр. вектор $x^i$).

2.2. При вычислении интеграла (2.4) необходимо учитывать, что в различных точках сферы $S$ величины $T^{ik}$ зависят от различных «точек излучения» на мировой линии частицы.

Пусть $z^i(\sigma)$ – мировая линия частицы. Определим вектор

(2.6) $$y^i(\sigma) = z^i(\sigma) - \gamma^i,$$

исходящий, как и вектор $x^i$ из 4-центра сферы $S$. Выберем на мировой линии частицы некоторую 4-точку $z^i$, и рассечем $S$ световыми конусами с вершиной $z^i$:

$$(x^i - y^i)\cdot(x_i - y_i) = 0$$

отсюда, учитывая (2.3), получаем уравнение 3-плоскости:

(2.7) $$x^i y_i = \frac{1}{2}\cdot y^i y_i$$

Пересечение 3-плоскости (2.7) со 2-сферой $S$ есть окружность $O$. Окружность $O$ представляет собой множество точек сферы $S$, для которых выбранная нами точка $z^i$ является «точкой излучения». Задача вычисления интеграла (2.4) сводится, таким образом, к вычислению интеграла по окружности $O$, а затем по 4-точкам $z^i$ вдоль мировой линии частицы.

Выберем на окружности $O$ точку $x^i$ и построим единичные 4-векторы $\eta^i$ и $\zeta^i$, ортогональные 4-векторам $x^i$ и $u^i$, причем так, что $\zeta^i$ параллелен плоскости (2.7), а $\eta^i$ – ортогонален $\zeta^i$. Легко убедиться, что этим условиям удовлетворяют 4-векторы

(2.8) $$\zeta^i = A_\zeta \cdot e^{iklm} \cdot x_k \cdot u_l \cdot y_m$$

(2.9) $$\eta^i = A_\eta \cdot e^{iklm} \cdot x_k \cdot u_l \cdot \zeta_m,$$

здесь $e^{iklm}$ – единичный, полностью антисимметричный псевдотензор, а нормировочные константы $A_\zeta$ и $A_\eta$ определим из условий:

(2.10) $$\zeta^i \cdot \zeta_i = -1, \qquad \eta^i \cdot \eta_i = -1.$$

При вычислении сверток (2.10) воспользуемся известным тождеством:

(2.11) $$e^{iklm} \cdot e_{ipqr} = -\delta_p^k \cdot \delta_q^l \cdot \delta_r^m - \delta_q^k \cdot \delta_r^l \cdot \delta_p^m - \delta_r^k \cdot \delta_p^l \cdot \delta_q^m + \delta_p^k \cdot \delta_r^l \cdot \delta_q^m + \delta_q^k \cdot \delta_p^l \cdot \delta_r^m + \delta_r^k \cdot \delta_q^l \cdot \delta_p^m$$

Поскольку мы считаем $y^i << R$, из (2.8)-(2.9) находим:

$$\text{(2.12)} \qquad \zeta^i = \frac{1}{R\sqrt{-y^p y_p}} \cdot e^{iklm} x_k u_l y_m$$

$$\text{(2.13)} \qquad \eta^i = \frac{1}{R} \cdot e^{iklm} x_k u_l \cdot \zeta_m ,$$

2.3. В качестве элемента интегрирования в (2.4) выберем прямоугольник, стороны которого параллельны векторам $\eta^i$ и $\zeta^i$.

Рассмотрим световой конус, вершина которого расположена в точке $\bar{y}^i$, лежащей на мировой линии частицы, достаточно близко к $y^i$, так, что можно считать

$$\text{(2.14)} \qquad \bar{y}^i = y^i + w^i \cdot \delta\sigma$$

где $w^i$ – вектор 4-скорости частицы. Подобно первому этот конус пересекается со сферой $S$ по окружности $\bar{O}$. Вычислим ширину полоски сферы $S$, заключенной между $O$ и $\bar{O}$. Для этого надо вычислить длину $l$ стороны прямоугольника, параллельную вектору $\eta^i$, заключенную между $O$ и $\bar{O}$. Поскольку конец вектора $x^i + l \cdot \eta^i$ лежит на световом конусе точки $\bar{y}^i$, можно составить уравнение

$$(x^i + l \cdot \eta^i - \bar{y}^i) \cdot (x_i + l \cdot \eta_i - \bar{y}_i) = 0$$

решая которое, находим:

$$\text{(2.15)} \qquad l = -\frac{w_i \cdot (x^i - y^i)}{\eta_k y^k} \cdot \delta\sigma$$

Скалярное произведение в знаменателе (2.15) легко вычисляется из (2.12) – (2.13):

$$\text{(2.16)} \qquad \eta^i y_i = \frac{1}{R} \cdot e^{iklm} x_k u_l \cdot \zeta_m y_i = -\frac{1}{R} \cdot \zeta_m \cdot e^{mkli} x_k u_l y_i = -\frac{1}{R} \cdot \zeta_m \cdot \zeta^m \cdot R\sqrt{-y_p y^p} = \sqrt{-y_p y^p}$$

С учетом (2.16) элемент интегрирования в (2.4) приобретает вид

$$\text{(2.17)} \qquad \delta S = \frac{(w_i \cdot (x^i - y^i)) \cdot \delta\sigma \cdot \delta q}{\sqrt{-y_p y^p}}$$

где $\delta q$ – сторона элементарного прямоугольника, параллельная вектору $\zeta^i$.

Теперь определим величину $\delta q$. Пусть $N^i$ – центр окружности $O$. Окружность $O$ является сечением сферы $S$ 2-плоскостью $a$, возникающей как пересечение 3-плоскости $t = R = const$ и 3-плоскости (2.7). Ясно, что вектор $N^i$ лежит в основании перпендикуляра, опущенного из центра сферы на 2–плоскость $a$. Отсюда следует, что вектор $N^i$ есть линейная комбинация векторов $u^i$ и $y^i$, удовлетворяющая уравнениям $N^i \cdot u_i = R$ и $N^i \cdot y_i = \frac{1}{2} y^j y_j$. Легко убедиться, что этим условиям удовлетворяет вектор

(2.18) $$N^i = \frac{1}{2} \cdot \frac{y^j y_j \cdot (u^j \cdot y_j - 2 \cdot R) \cdot u^i + (2 \cdot R \cdot u^m y_m - y_m y^m) \cdot y^i}{(u^m y_m)^2 - y^m y_m}$$

Рассмотрим вектор $\psi^i$, из центра окружности в какую-либо точку на $O$. Поскольку $\psi^i$ лежит в плоскости $t = const$, то $\psi^0 = 0$ и радиус окружности $O$ $R_o^2 = -\psi_k \psi^k$. С другой стороны конец вектора $\psi^i$ лежит на сфере $S$, откуда следует:

(2.19) $$(N^i + \psi^i) \cdot (N_i + \psi_i) = 0$$

Таким образом, мы приходим к выражению для радиуса окружности $O$:

(2.20) $$R_o^2 = N^i \cdot N_i = -\frac{y_k y^k}{4} \cdot \frac{4 \cdot R^2 - 4 \cdot R \cdot u^m y_m + y^k \cdot y_k}{(u^m y_m)^2 - y_m y^m}$$

Введем в 3-пространстве неподвижного относительно сферы $S$ наблюдателя систему сферических координат $(\theta, \varphi)$ с центром в центре $S$ и полярной осью, направленной вдоль пространственных компонент вектора $y^i$. Иными словами, полярная ось направлена из центра сферы в выбранную «точку излучения» $z^i$.

Как видно из (2.18), в лабораторной системе (где исчезают пространственные компоненты вектора $u^i$) 3–векторы $y^\alpha$ и $N^\alpha$ коллинеарны, и, следовательно, полярная ось перпендикулярна плоскости контура $O$.

В этой системе элемент интегрирования вдоль $O$ можно записать как

(2.21) $$dq = R_o \cdot d\varphi$$

Выражение в знаменателе (2.20), как легко видеть, есть квадрат 3-расстояния от центра сферы до «точки излучения» в лабораторной системе. Обозначим его через $r$:

(2.22) $$r = \sqrt{(u^m y_m)^2 - y_m y^m}.$$

Учитывая наше предположение $y^i << R$, преобразуем (2.20):

(2.23) $$R_o = \frac{\sqrt{-y_k y^k}}{r} \cdot R$$

Все векторы $N^i + \psi^i$ из центра сферы к произвольной точке окружности $O$ образуют с 3-вектором $y^\alpha$ одинаковый угол, который мы обозначим $\theta_s$. Косинус угла $\theta_s$ вычислим как скалярное произведение 3-векторов $N^\alpha + \psi^\alpha$ и $y^\alpha$, деленное на произведение длин соответствующих 3-векторов:

$$\text{(2.24)} \qquad cos\theta_s = \frac{(N^i+\psi^i)\cdot u_i\cdot y^k\cdot u_k-(N^m+\psi^m)\,y_m}{r\cdot R}=\frac{y^k\cdot u_k}{r}-\frac{y^k\,y_k}{2\cdot r\cdot R}$$

или, пренебрегая слагаемым порядка $R^{-1}$ :

$$\text{(2.25)} \qquad cos\theta_s = \frac{y^k\cdot u_k}{r}$$

Из (2.17), (2.21) и (2.23) найдем элемент интегрирования по сфере **S**:

$$\text{(2.26)} \qquad \delta\mathbf{S} = \frac{R\cdot(w_i\cdot(x^i-y^i))\cdot\delta\sigma\cdot\delta\varphi}{r}$$

и теперь (2.4) мы можем привести к выражению:

$$\text{(2.27)} \qquad \frac{dP^i}{dt} = \int_{\sigma_1}^{\sigma_2} d\sigma \int_0^{2\cdot\pi} d\varphi\cdot T^{ik}\cdot n_k \cdot \frac{R}{r}\cdot(w_m\cdot(x^m-y^m))$$

Пределы интегрирования $\sigma_1$, $\sigma_2$ определим из условия, что конец вектора $N^i$ оказывается на поверхности сферы **S**. Это приводит к уравнению относительно $y^i(\sigma)$.

$$\text{(2.28)} \qquad N^i\cdot N_i = 0$$

Согласно (2.20) эта свертка распадается на два сомножителя, второй из которых всегда положителен, так как мы предположили, что радиус сферы $R$ достаточно велик. Таким образом, пределы $\sigma_1$, $\sigma_2$ определяются уравнением

$$\text{(2.29)} \qquad y^i(\sigma)\cdot y_i(\sigma) = 0$$

Уравнение (2.29) имеет ясный геометрический смысл (см. рис. 1). Луч света из 4-точки $y^i(\sigma_1)$ (луч AB) достигает самой удаленной от частицы точки сферы **S**. Тогда как луч света из точки $y_i(\sigma_2)$ (луч CD) достигает сферу **S** в ближайшей к частице точке. Оба луча AB и CD лежат на световом конусе с вершиной в центре сферы **S**. Таким образом, отрезок интегрирования в (2.19) [$\sigma_1$, $\sigma_2$] вырезается на мировой линии частицы световым конусом центра сферы **S**, причем $y^i(\sigma_1)$ лежит на границе конуса абсолютного прошлого, а $y^i(\sigma_2)$ — абсолютного будущего. Все же «точки излучения», создающие поток энергии-импульса через поверхность сферы **S** отделены от 4-центра сферы $y^i$ пространственноподобными интервалами, так что в интервале ($\sigma_1$, $\sigma_2$) имеет место:

$$\text{(2.30)} \qquad y^i(\sigma)\cdot y_i(\sigma) < 0, \qquad \sigma \in (\sigma_1,\ \sigma_2)$$

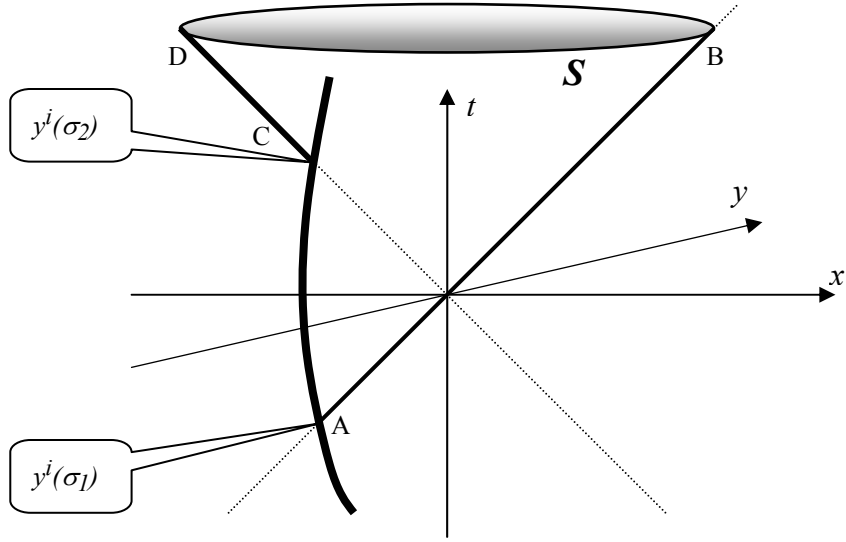

Рис. 1. Мировая линии частицы (АС) и удаленная неподвижная сфера *S*, с центром в 4-точке $\gamma^i$. «Точки излучения» для всей сферы *S* расположены на участке АС.

### 3. Преобразование тензора энергии-импульса.

Вычислим подынтегральное выражение в (2.4) с помощью формул (1.2) и (1.3). При этом будем оставлять только члены, убывающие с расстоянием как $R^{-2}$. Во-первых заметим, что свертка 4-тензора $F^{ik}$ с 4-вектором запаздывания $\chi^i$ убывает как $R^{-1}$:

(3.1) $$h^i = F^{ik} \cdot \chi_k = \frac{e}{(\chi^j \cdot w_j)^2} \cdot \chi^i \sim R^{-1}$$

Опустив оба индекса в (1.2) и свернув затем с $F^{ik}$, используя (3.1) найдем:

(3.2) $$F^{ik} \cdot F_{ik} = \frac{e}{(\chi^j \cdot w_j)^3} \cdot (-h^k \cdot w_k - w_i \cdot h^i) \cdot (1 - \chi^m a_m) + \frac{e}{(\chi^j \cdot w_j)^2} \cdot (-h^k \cdot a_k - a_i \cdot h^i)$$

или

(3.3) $$F^{ik} \cdot F_{ik} = -\frac{2 \cdot e^2}{(\chi^j \cdot w_j)^4} \sim R^{-4}.$$

Мы видим, что второе слагаемое в (1.3) дает исчезающе малый вклад в поток энергии-импульса на удаленной сфере.

Рассмотрим первое слагаемое в (1.3).

(3.4) $$F^{ik} \cdot F_{lk} \cdot n^l = -\frac{1}{R} \cdot F^{ik} \cdot F_{lk} \cdot x^l + \frac{1}{R} \cdot F^{ik} \cdot F_{0k} \cdot x^0 = -\frac{1}{R} \cdot F^{ik} \cdot F_{lk} \cdot \chi^l - \frac{1}{R} \cdot F^{ik} \cdot F_{lk} \cdot y^l + F^{ik} \cdot F_{0k}$$

Первое слагаемое в (3.4) мало в силу (3.1), второе – из-за $R >> y^i$, и только третье включает члены порядка $R^{-2}$. Опустив индексы в (1.2) и исключив слагаемые $\sim R^{-3}$, найдем:

(3.5) $$F^{ik} \cdot F_{lk} \cdot n^l = F^{ik} \left( \frac{e}{(\chi^j \cdot w_j)^3} \cdot w_k \cdot \chi_0 \cdot (-\chi^m a_m) + \frac{e}{(\chi^j \cdot w_j)^2} \cdot a_k \cdot \chi_0 \right)$$

Учитывая, что $\chi_0 = R$ из (1.2) находим после упрощений:

(3.6) $$F^{ik} \cdot F_{lk} \cdot n^l = \frac{e^2 \cdot R}{(\chi^j \cdot w_j)^6} \cdot \chi^i \cdot (\chi^m a_m)^2 + \frac{e^2 \cdot R}{(\chi^j \cdot w_j)^4} \cdot \chi^i \cdot a^k \cdot a_k$$

Поскольку $x^i - \chi^i = y^i << R$ с точностью до членов, исчезающих при больших $R$, мы можем просто заменить в (3.6) вектор запаздывания $\chi^i$ на вектор $x^i$ – из центра сферы $S$ к поверхности интегрирования. Таким образом, (2.27) приводится к следующему выражению для мощности излучения:

(3.7) $$\frac{dP^i}{dt} = -\frac{1}{4\pi} \cdot \int_{\sigma_1}^{\sigma_2} d\sigma \int_0^{2\pi} d\varphi \cdot \frac{e^2 \cdot R^2 \cdot x^i}{r \cdot (x^n \cdot w_n)^5} \cdot \left( (x^m a_m)^2 + (x^n \cdot w_n)^2 \cdot a^k \cdot a_k \right)$$

При вычислении интегралов (3.7) при каждом значении параметра $\sigma$ выбирается полярная ось из центра сферы в направлении на текущее положение частицы, затем вычисляется интеграл по углу $\varphi$ при фиксированном значении угла $\theta_s$, определяемом в (2.25).

Определим следующий интеграл:

(3.8) $$K^{ikm}(\theta) = \int_0^{2\pi} \frac{x^i \cdot x^k \cdot x^m \cdot d\varphi}{(w^j \cdot x_j)^5}$$

Интеграл (3.8) вычисляется по окружности, расположенной на поверхности единичной сферы, и такой, что все ее точки видны из центра сферы под углом $\theta$ к полярной оси.

С помощью (3.8) интеграл (3.7) приводится к виду:

(3.9) $$\frac{dP^i}{dt} = -\frac{e^2}{4\pi} \cdot \int_{\sigma_1}^{\sigma_2} \frac{d\sigma}{r} \cdot \left( K^{ikm} \cdot a_k \cdot a_m + K^{ikm} w_k \cdot w_m \cdot a^n \cdot a_n \right)$$

На следующем шаге вычислим интеграл (3.8).

## 4. Вычисление интеграла по азимутальному углу.

4.1. Введём вспомогательный произвольный 4-вектор $\zeta^i$ и определим функцию $\Theta$:

$$\Theta = \frac{1}{(\zeta_k \cdot x^k)^2} \qquad (4.1)$$

Легко видеть, что подынтегральное выражение в (3.8) выражается через производные $\Theta$:

$$\frac{x^i x^k x^m}{(w^j \cdot x_j)^5} = -\frac{1}{24} \frac{\partial}{\partial \zeta_i} \frac{\partial}{\partial \zeta_k} \frac{\partial}{\partial \zeta_m} \Theta \bigg|_{\zeta_j = w_j} \qquad (4.2)$$

где после дифференцирования вектор $\zeta_j$ должен быть заменён вектором 4-скорости $w_j$. Поскольку дифференцирование по нефизическому параметру и интегрирование по азимутальному углу коммутируют, то для упрощения рассуждений вычислим сначала интеграл функции $\Theta$, а затем дифференцированием получим выражение для $K^{ikm}$. Вектор $x^i$, входящий в (4.1) имеет компоненты:

$$x^i = \{1,\ \sin\theta \cdot \cos\varphi,\ \sin\theta \cdot \sin\varphi,\ \cos\theta\} \qquad (4.3)$$

а свёртка в знаменателе (4.1):

$$\zeta_k \cdot x^k = \zeta_0 - \zeta_1 \cdot \sin\theta \cdot \cos\varphi - \zeta_2 \cdot \sin\theta \cdot \sin\varphi - \zeta_3 \cdot \cos\theta \qquad (4.4)$$

Введём обозначения:

$$a = \zeta_0 - \zeta_3 \cdot \cos\theta\ ;\qquad b = \sin\theta \sqrt{\zeta_1^2 + \zeta_2^2}\ ;\qquad \cos\varphi_m = \frac{\zeta_1}{\sqrt{\zeta_1^2 + \zeta_2^2}} \qquad (4.5)$$

тогда (4.4) преобразуется к виду:

$$\zeta_k \cdot x^k = a - b \cdot \cos(\varphi - \varphi_m)\ . \qquad (4.6)$$

Заметим, что если предполагать вектор $\zeta_k$ мало отличающимся от 4-вектора скорости $w_k$, то его квадрат заведомо положителен, откуда следует соотношение

$$a > |b|\ , \qquad (4.7)$$

которое будет использовано ниже. Интеграл от функции $\Theta$ представим в виде:

$$\int_0^{2\pi} \Theta \cdot d\varphi = \int_0^{2\pi} \frac{d\varphi}{(a - b \cdot \cos\varphi)^2} \qquad (4.8)$$

так как в силу периодичности косинуса замена $\varphi - \varphi_m \to \varphi$ не меняет результата.

4.2. С помощью замены переменной

$$\xi = tg(\varphi/2) \qquad (4.9)$$

преобразуем интеграл (4.8) к виду:

$$\text{(4.10)} \quad \int_0^{2\pi} \Theta \cdot d\varphi = 2 \cdot \int_{-\infty}^{\infty} \frac{(1+\xi^2)}{(a-b+\xi^2 \cdot (a+b))^2} d\xi$$

Будем считать $\xi$ в (4.10) комплексной переменной. Поскольку при больших по модулю $\xi$ подынтегральное выражение пропорционально $\sim \xi^{-2}$, то интеграл по удаленной дуге, огибающей верхнюю полуплоскость, мал. Следовательно, интеграл (4.10) сводится к интегралу по контуру, обходящему единственный полюс $\xi_0$ в верхней полуплоскости:

$$\text{(4.11)} \quad \xi_0 = i \cdot \sqrt{\frac{a-b}{a+b}}$$

(из неравенства (4.7) следует, что $\xi_0$ расположен на мнимой оси в положительной части). Сделав в (4.10) замену переменной

$$\xi = \xi_0 \cdot (1 + 2 \cdot x)$$

придем к выражению

$$\text{(4.12)} \quad \int_0^{2\pi} \Theta \cdot d\varphi = \frac{1}{4 \cdot \xi_0^3 \cdot (a+b)^2} \cdot \int_C \frac{1 + \xi_0^2 \cdot (1 + 2 \cdot x)^2}{x^2 \cdot (x+1)^2} dx$$

где контур $C$ обходит начало координат против часовой стрелки. По теореме о вычетах:

$$\text{(4.13)} \quad \int_0^{2\pi} \Theta \cdot d\varphi = \frac{2 \cdot \pi \cdot i}{4 \cdot \xi_0^3 \cdot (a+b)^2} \cdot (4 \cdot \xi_0^2 - 2 \cdot (1 + \xi_0^2))$$

Подставляя (4.11) и упрощая, получим:

$$\text{(4.14)} \quad \int_0^{2\pi} \Theta \cdot d\varphi = \frac{2 \cdot \pi \cdot a}{(a^2 - b^2)^{3/2}}$$

4.3. Выразим величины $a$ и $a^2 - b^2$ через данные нам 4-векторы $\zeta^i$, $y^i$, $u^i$. 4-вектор $y^i$ в рассматриваемой системе координат имеет компоненты:

$$\text{(4.15)} \quad y^i = \{y^0, 0, 0, r\}$$

Очевидны следующие равенства

$$\text{(4.16)} \quad \zeta_0 = \zeta_k u^k \; ; \quad \zeta_i y^i = \zeta_0 y^0 - \zeta_3 \cdot r \; ; \quad \zeta_1^2 + \zeta_2^2 = -(\zeta^m \cdot \zeta_m - \zeta_0^2 + \zeta_3^2)$$

Используя (4.16), (2.25) и (2.22) находим из (4.5) после элементарных преобразований:

$$\text{(4.17)} \quad a = \zeta_k \cdot \frac{-y^m y_m \cdot u^k + y^k \cdot y^m \cdot u_m}{r^2}$$

$$(4.18) \quad a^2-b^2 = \frac{(\zeta_k \cdot y^k)^2 - y^k \cdot y_k \cdot \zeta^m \cdot \zeta_m}{r^2}$$

Нам удобно ввести вспомогательный симметричный тензор

$$(4.19) \quad h^{ik} = (y^i \cdot y^k - g^{ik} \cdot y^m \cdot y_m)$$

и выразить с его помощью величины $a$ и $a^2-b^2$:

$$(4.20) \quad a = \frac{h^{ik} \cdot u_i \cdot \zeta_k}{r^2}$$

$$(4.21) \quad a^2-b^2 = \frac{h^{ik} \cdot \zeta_i \cdot \zeta_k}{r^2}$$

Таким образом, для искомого интеграла из (4.14) находим выражение

$$(4.22) \quad \int_0^{2\pi} \Theta \cdot d\varphi = \frac{2\pi \cdot r \cdot h^{ik} \cdot u_i \cdot \zeta_k}{(h^{ik} \cdot \zeta_i \cdot \zeta_k)^{3/2}}$$

Для величин $K^{ikm}(\theta)$ из (3.8) находим:

$$(4.23) \quad K^{ikm}(\theta) = -\frac{1}{24} \frac{\partial}{\partial \zeta_i} \frac{\partial}{\partial \zeta_k} \frac{\partial}{\partial \zeta_m} \frac{2\pi \cdot r \cdot h^{pq} \cdot u_p \cdot \zeta_q}{(h^{pq} \cdot \zeta_p \cdot \zeta_q)^{3/2}} \bigg|_{\zeta_j = w_j}$$

А для мощности излучения из (3.9):

$$(4.24) \quad \frac{dP^i}{dt} = \frac{e^2}{48} \cdot \int_{\sigma_1}^{\sigma_2} d\sigma \cdot (a_k \cdot a_m + w_k \cdot w_m \cdot a^n \cdot a_n) \cdot \frac{\partial}{\partial \zeta_i} \frac{\partial}{\partial \zeta_k} \frac{\partial}{\partial \zeta_m} \frac{h^{pq} \cdot u_p \cdot \zeta_q}{(h^{pq} \cdot \zeta_p \cdot \zeta_q)^{3/2}} \bigg|_{\zeta_j = w_j}$$

Выполнив дифференцирование в (4.24), получим:

$$(4.25) \quad \frac{dP^i}{dt} = \frac{e^2}{16} \cdot \int_{\sigma_1}^{\sigma_2} d\sigma \cdot (a_k \cdot a_m + w_k \cdot w_m \cdot a^n \cdot a_n) \cdot Q^{ikm},$$

$$(4.26) \quad Q^{ikm} = -\frac{\lambda^m \cdot h^{ik} + \lambda^i \cdot h^{km} + \lambda^k \cdot h^{im}}{(h^{pq} \cdot w_p \cdot w_q)^{5/2}} + 5 \cdot \frac{\lambda^m \cdot \mu^k \cdot \mu^i + \lambda^k \cdot \mu^m \cdot \mu^i + \lambda^i \cdot \mu^m \cdot \mu^k}{(h^{pq} \cdot w_p \cdot w_q)^{7/2}} + 5 \cdot \frac{\lambda^q \cdot w_q \cdot (h^{km} \cdot \mu^i + h^{im} \cdot \mu^k + \mu^m \cdot h^{ik})}{(h^{pq} \cdot w_p \cdot w_q)^{7/2}} - 35 \cdot \frac{\lambda^q \cdot w_q \cdot \mu^m \cdot \mu^k \cdot \mu^i}{(h^{pq} \cdot w_p \cdot w_q)^{9/2}}$$

где введены обозначения:

$$(4.27) \quad \lambda^i = h^{ip} \cdot u_p; \qquad \mu^i = h^{ip} \cdot w_p;$$

Выражение (4.24) может быть удобно представить через четырехкратное дифференцирование по нефизическому параметру $\zeta_i$

$$(4.28) \quad \frac{dP^i}{dt} = -\frac{e^2}{48} \cdot \int_{\sigma_1}^{\sigma_2} d\sigma \cdot (a_k \cdot a_m + w_k \cdot w_m \cdot a^n \cdot a_n) \cdot u^l \cdot \frac{\partial}{\partial \zeta_i} \frac{\partial}{\partial \zeta_k} \frac{\partial}{\partial \zeta_m} \frac{\partial}{\partial \zeta_l} \frac{1}{(h^{pq} \cdot \zeta_p \cdot \zeta_q)^{1/2}} \bigg|_{\zeta_j = w_j}$$

## 5. Вычисление потери энергии частицей в выбранной точке мировой линии

Теперь мы можем поставить вопрос о радиационных потерях энергии частицы в некоторой точке. Для этого выберем центр сферы ***S***, которую до сих пор мы считали произвольной, вблизи рассматриваемой точки мировой линии.

В этом случае интервал интегрирования в (4.25) или (4.28) $\delta\sigma = \sigma_2 - \sigma_1$ (участок AS на рис. 1) окажется малым. Как легко усмотреть из рис. 1, выбирая центр сферы достаточно близко к мировой линии, мы можем сделать величину $\delta\sigma$ сколь угодно малой. С точностью до членов, не исчезающих при $\delta\sigma \to 0$, мы можем считать величины $w^i$ и $a^i$ постоянными на участке интегрирования и вынести их за знак интеграла.

Вектор $y^i$ имеет первый порядок малости по $\delta\sigma$ и может быть представлен в виде:

$$(5.1) \quad y^i = \beta^i + w^i \sigma,$$

где $\beta^i = y^i(\sigma_1)$ -- 4-вектор из центра сферы ***S*** в точку A (см. рис. 1) – лежащий на световом конусе абсолютного прошлого конец интервала интегрирования. (В выражении (5.1) предполагается, что $\sigma_1 = 0$):

$$(5.2) \quad \beta^i \beta_i = 0$$

Величину верхнего предела интегрирования в (4.25) $\sigma_2$ найдем из условия, что вектор $y(\sigma_2)$ светоподобен. Из (5.1) находим:

$$y^i(\sigma_2) y_i(\sigma_2) = (\beta^i + w^i \sigma_2)(\beta_i + w_i \sigma_2) = \sigma_2 \cdot (\sigma_2 + 2 \cdot w^i \beta_i) = 0,$$

откуда видно, что

$$(5.3) \quad \sigma_2 = -2 w^i \beta_i$$

Обозначим через $U$ свертку векторов $w^i$ и $\beta^i$:

$$(5.4) \quad U = -w^i \beta_i$$

тогда

$$(5.5) \quad \sigma_2 = 2U.$$

Поскольку вектор $\beta^i$ лежит на световом конусе абсолютного прошлого, ясно что

$$(5.6) \quad U > 0.$$

Из (4.19),(5.1),(5.2) вычислим свертку $h^{ik}w_i w_k$, входящую в знаменатель (4.26):

(5.7) $$h^{ik}w_i w_k = (\beta^i + w^i\sigma)w_i(\beta^k + w^k\sigma)w_k - (\beta^i + w^i\sigma)(\beta_i + w_i\sigma) = U^2$$

Из (5.7) видно, что в рассматриваемом приближении свертка $h^{ik}w_i w_k$, входящая в знаменатель (4.26), постоянна на отрезке интегрирования, в связи с чем выражение (4.26) есть полином степени относительно $\sigma$. Интегрирование (4.26) в интервале $\sigma \in (0, 2U)$ – несколько трудоемкая, но по существу тривиальная выкладка. Это вычисление содержится в Приложении 1, где получен следующий результат:

(5.8) $$\frac{dP^i}{dt} = e^2 \cdot \left( -\frac{14}{15} w^i w^p \cdot u_p \cdot a^n a_n - \frac{2}{15} \cdot a^i \cdot a^p u_p + \frac{4}{15} \cdot u^i \cdot a^n a_n \right)$$

**6. Прямолинейное движение заряженной частицы.**

В случае прямолинейного движения можно существенно упростить выражение для мощности потерь (4.25)–(4.26). Выберем центр сферы на 3-прямой, вдоль которой движется частица, и направим ось $z$ по этой прямой. При этом мы не требуем, чтобы 4-центр сферы $S$ располагался на мировой линии частицы. Тогда компоненты вектора $y^i$:

(6.1) $$y^i = \{y^0; 0; 0; y^3\}$$

А тензор $h^{ik}$ из (4.19):

(6.2) $$h^{ik} = \begin{vmatrix} y^3 y^3 & 0 & 0 & y^3 y^0 \\ 0 & y^0 y^0 - y^3 y^3 & 0 & 0 \\ 0 & 0 & y^0 y^0 - y^3 y^3 & 0 \\ y^3 y^0 & 0 & 0 & y^0 y^0 \end{vmatrix}$$

Из (6.2) видно, что тензор $h^{ik}$ можно представить в виде:

(6.3) $$h^{ik} = p^i p^k + \pi^{ik}$$

где

(6.4) $$p^i = \{y^3; 0; 0; y^0\}$$

(6.5) $$\pi^{ik} = \begin{vmatrix} 0 & 0 & 0 & 0 \\ 0 & y^0 y^0 - y^3 y^3 & 0 & 0 \\ 0 & 0 & y^0 y^0 - y^3 y^3 & 0 \\ 0 & 0 & 0 & 0 \end{vmatrix}$$

Из (4.25)-(4.27) видно, что тензор $h^{ik}$ входит в выражение для потерь энергии-импульса только в составе свертки с одним из векторов $w^i, a^i$ или $u^i$. А поскольку компоненты этих векторов вдоль осей *x* и *y* равны нолю, то равны нолю и свертки этих векторов с тензором $\pi^{ik}$. Перепишем выражение (4.26), опуская вклад тензора $\pi^{ik}$:

$$(6.6) \qquad Q^{ikm} = -\frac{8 \cdot p^i p^k p^m p^q}{\left|p^n w_n\right|^5} u_q$$

Откуда для потерь энергии получаем выражение:

$$(6.7) \qquad \frac{dP^i}{dt} = -\frac{e^2}{2} \cdot \int_{\sigma_1}^{\sigma_2} d\sigma \cdot \left(a_k a_m + w_k w_m \cdot a^l a_l\right) \cdot \frac{p^i p^k p^m p^q}{\left|p^n w_n\right|^5} u_q$$

Бывает удобно перейти от интегрирования по $d\sigma$ к переменной

$$(6.8) \qquad \xi = \frac{y^0}{y^3}$$

Согласно (2.25), величина $\xi$ представляет собой косинус угла наклона наблюдаемого излучения к оси *z*. Дифференцируя (6.8) по $d\sigma$, получаем:

$$(6.9) \qquad d\sigma = \frac{y^3}{w^0 - \xi \cdot w^3} d\xi = \frac{y^3}{w^k v_k} d\xi$$

Где через $v^k$ обозначен единичный 4-вектор из центра сферы в точку наблюдения (если считать $\xi = \cos\vartheta$, где $\vartheta$ – полярный угол точки наблюдения):

$$(6.10) \qquad v^i = \{1; \sin\vartheta \cdot \cos\varphi; \sin\vartheta \cdot \sin\varphi; \cos\vartheta\}$$

Учитывая (6.8)-(6.10), формулу (6.7) перепишем в виде:

$$(6.11) \qquad \frac{dP^i}{dt} = -\frac{e^2}{2} \cdot \int_{-1}^{1} d\xi \cdot \left(a_k a_m + w_k w_m \cdot a^l a_l\right) \cdot \frac{v^i v^k v^m v^q}{\left(v^n w_n\right)^6} u_q$$

Формула (6.11) определяет угловое распределение излучения.

## 7. Гиперболическое движение заряженной частицы.

Гиперболическим называют движение частицы, при котором она испытывает постоянное ускорение в сопутствующей системе отсчета. Обозначив величину этого ускорения $\lambda$ и выбрав надлежащим образом систему координат, уравнение мировой линии частицы можно записать в виде:

$$(7.1) \qquad z^i = \{\lambda^{-1} \cdot sh(\lambda\sigma); 0; 0; \lambda^{-1} \cdot ch(\lambda\sigma)\}$$

где $\sigma$ – естественная параметризация мировой линии или «собственное время» частицы.

Гиперболическое движение часто называют движением «равноускоренного заряда», однако этот термин нам представляется менее удачным, так как в системе отсчета неподвижного наблюдателя (в лабораторной системе) только на нерелятивистском участке частица движется равноускоренно.

Пусть центр сферы **S** расположен на прямой, вдоль которой движется частица. Обозначим $s$ и $d$ его 4-координаты так, что 4-вектор $\gamma^i$ примет вид:

(7.2) $$\gamma^i = \{s; 0; 0; d;\}$$

Тогда для вектора $y^i$ получим:

(7.3) $$y^i = \{\lambda^{-1} \cdot sh(\lambda\sigma) - s; 0; 0; \lambda^{-1} \cdot ch(\lambda\sigma) - d\}$$

Дифференцируя (6.1), для векторов $w^i$, $a^i$ получим выражения:

(7.4) $$w^i = \{ch(\lambda\sigma); 0; 0; sh(\lambda\sigma)\} \quad a^i = \{\lambda \cdot sh(\lambda\sigma); 0; 0; \lambda \cdot ch(\lambda\sigma)\}$$

Подставив эти значения в (6.11), находим:

(7.5) $$\frac{dP^i}{dt} = \frac{e^2 \lambda^2}{2} \cdot \int_{-1}^{1} d\xi \cdot \frac{v^i \cdot (1-\xi^2)}{\left(ch(\lambda\sigma) - \xi \cdot sh(\lambda\sigma)\right)^6}$$

Величина $\sigma$ в знаменателе (6.5) связана с $\xi$ неявной зависимостью, которую мы найдем из (6.8) и (7.3):

(7.6) $$\xi = \frac{\lambda^{-1} \cdot sh(\lambda\sigma) - s}{\lambda^{-1} \cdot ch(\lambda\sigma) - d}$$

Отсюда находим уравнение для $\sigma$:

(7.7) $$\xi \cdot ch(\lambda \cdot \sigma) - m = sh(\lambda \cdot \sigma)$$

где введено обозначение

(7.8) $$m = \lambda \cdot (d \cdot \xi - s)$$

Обозначим $q = ch(\lambda \cdot \sigma)$ и, возведя в квадрат (7.7), придем к уравнению для $q$:

(7.9) $$(1-\xi^2) \cdot q^2 + 2 \cdot m \cdot \xi \cdot q - (1+m^2) = 0$$

Положительный корень (7.9) определяется выражением:

(7.10) $$q = \frac{-m \cdot \xi + \sqrt{m^2 + 1 - \xi^2}}{1 - \xi^2}$$

Теперь из (7.7) и (7.10) находим выражение для знаменателя (7.5):

(7.11) $$ch(\lambda\sigma) - \xi \cdot sh(\lambda\sigma) = q \cdot (1-\xi^2) + \xi \cdot m = \sqrt{m^2 + 1 - \xi^2}$$

Из (7.5), (7.8) и (7.11) приходим к следующему выражению для излучения:

(7.12) $$\frac{dP^i}{dt} = \frac{e^2 \lambda^2}{2} \cdot \int_{-1}^{1} d\xi \cdot \frac{v^i \cdot (1-\xi^2)}{\left(\lambda \cdot (d \cdot \xi - s)^2 + 1 - \xi^2\right)^3}$$

Вычисление интеграла (7.12) слегка упрощается, если дополнительно предположить, что центр сферы **S** расположен на мировой линии частицы (7.1), в 4-точке с некоторым значением параметра $\sigma=\psi$:

(7.13) $$s = \lambda^{-1} \cdot sh(\lambda\psi); \quad d = \lambda^{-1} \cdot ch(\lambda\psi);$$

Тогда из (7.12) и (7.13) легко найти:

(7.14) $$\frac{dP^i}{dt} = \frac{e^2 \lambda^2}{2} \cdot \int_{-1}^{1} d\xi \cdot \frac{v^i \cdot (1-\xi^2)}{\left(ch(\lambda\psi) - \xi \cdot sh(\lambda\psi)\right)^6}$$

Из подынтегрального выражения в (7.14) легко получить угловое распределение энергии излучения, если заменить $\xi=cos\theta$. Интегралы (7.14) могут быть вычислены интегрированием по частям. В результате находим:

(7.15) $$\frac{dP^0}{dt} = \frac{2 \cdot e^2 \cdot \lambda^2}{15} \cdot (sh^2(\lambda\psi) + 5 \cdot ch^2(\lambda\psi))$$

(7.16) $$\frac{dP^0}{dt} = \frac{4 \cdot e^2 \cdot \lambda^2}{5} \cdot ch(\lambda\psi) \cdot sh(\lambda\psi)$$

Поскольку мы выбрали 4-центр сферы на мировой линии частицы в точке $\sigma=\psi$, то ясно, что весь поток энергии на сфере **S** соответствует «точке излучения» $\sigma=\psi$. При этом из (7.15) видно, что если точка $\psi$ выбрана на нерелятивистском участке мировой линии (то есть $\lambda\psi<<1$), то (7.15) сводится к формуле Лармора:

(7.17) $$\frac{dP^0}{dt} = \frac{2 \cdot e^2 \cdot \lambda^2}{3}$$

Однако если выбрать $\psi$ на ультрарелятивистском участке (где $ch(\lambda\psi)>>1$), то безразмерный коэффициент становится иным:

(7.18) $$\frac{dP^0}{dt} = \frac{4 \cdot e^2 \cdot \lambda^2}{5} \cdot \gamma^2$$

где $\gamma=ch(\lambda\cdot\psi)=\frac{K}{m}$ – релятивистский фактор – отношение кинетической энергии частицы к массе покоя. Соотношения (7.17) и (7.18) могут быть получены непосредственно из (5.8).

Ранее, в работе [3], мы получили эквивалентные (7.14)-(7.18) выражения в 3-геометрии. Там же предложено физическое осмысление этих результатов.

## 8. Магнитотормозное излучение.

Магнитотормозным в литературе называют излучение частицы, движущейся по винтовой линии в магнитном поле. Магнитотормозное излучение играет важнейшую роль при изучении пульсаров, внегалактических радиоисточников и иных астрофизических явлений. Для данной теоретической работы вопрос о магнитотормозном излучении имеет принципиальное значение, поскольку на этом примере Гинзбургом, Сазоновым и Сыроватским [1] в середине 60-х годов был решен вопрос о корректной методике вычисления потерь энергии движущейся частицей.

Мировая линии частицы, движущейся в магнитном поле, определяется выражением:

(8.1) $$z^i = \{t; \rho \cdot \cos(\omega t); \rho \cdot \sin(\omega t); V \cdot t\}$$

здесь ось $z$ направлена вдоль силовой линии магнитного поля, $\rho$ – постоянный радиус окружности, а $\omega$ – циклическая частота (в лабораторной системе).

8.1. В [1] вычислена напряженность в волновой зоне электрического поля, возникающего при движении по мировой линии (8.1). Компонента Фурье с номером $n$ выражается формулой:

(8.2) $$\vec{E}_n = \frac{2e}{r} \cdot \frac{n\rho\omega^2}{(1-Vk_z)^2} \left\{ \vec{l}_1 \cdot J'_n(z_n) - i\vec{l}_2 \cdot \frac{k_z - V}{\rho\omega \cdot k_\perp} \cdot J_n(z_n) \right\}$$

где

(8.3) $$z_n = n \cdot \frac{\rho\omega \cdot k_\perp}{1-Vk_z}$$

$k_z, k_\perp$ – компоненты единичного вектора $\vec{k}$, направленного на точку наблюдения;

$\vec{l}_1, \vec{l}_2$ – единичные векторы, ортогональные $\vec{k}$; $J_n(z)$ – функция Бесселя.

$r$ – расстояние до точки наблюдения.

Средняя за период мощность излучения частицы на частоте $n$-ой гармоники:

(8.4) $$\overline{S}_n = \frac{1}{8\pi} \cdot \vec{E}_n \cdot \vec{E}_n^*$$

Полная средняя потеря энергии частицы вычисляется суммированием величин $S_n$ по гармоникам и интегрированием по поверхности удаленной сферы. Это вычисление выполнено в Приложении 2. Вычисления приводят к следующему результату:

(8.5) $$\frac{d\overline{P}^0}{dt} = \frac{e^2\rho^2\omega^4}{1-V^2} \cdot \left( \frac{2}{3}\gamma^4 + \frac{V^2\gamma^2}{1-V^2} \cdot \Psi(\xi) \right)$$

где введено обозначение

$$\xi = \frac{\rho\omega}{\sqrt{1-V^2}}, \quad (8.6)$$

а $\Psi(\xi)$ – следующая функция одной переменной, определенная на интервале $(0,1)$:

$$\Psi(\xi) = \frac{1}{16} \cdot \left( \frac{2}{\xi^4} + \frac{14}{3\xi^2} - \left( \frac{3}{\xi} - \frac{2}{\xi^3} - \frac{1}{\xi^5} \right) \cdot \ln\left|\frac{1-\xi}{1+\xi}\right| \right) \quad (8.7)$$

Функция $\Psi(\xi)$ монотонно возрастает от $\frac{4}{15}$ (при $\xi = 0$) к $\frac{5}{12}$ (при $\xi = 1$).

8.2. «Собственное время» частицы $d\sigma$ связано с лабораторным временем через релятивистский фактор $\gamma$:

$$dt = \gamma \cdot d\sigma = \frac{d\sigma}{\sqrt{1 - \rho^2\omega^2 V^2}} \quad (8.8)$$

Дифференцируя (8.1) по $d\sigma$, получим выражения для 4-скорости и 4-ускорения:

$$w^i = \frac{dz^i}{d\sigma} = \gamma \cdot \{1; -\rho\cdot\omega\cdot\sin(\omega t); \rho\cdot\omega\cdot\cos(\omega t); V\} \quad (8.9)$$

$$a^i = \frac{dw^i}{d\sigma} = -\gamma^2\omega^2 \cdot \{0; \rho\cdot\cos(\omega t); \rho\cdot\sin(\omega t); 0\} \quad (8.10)$$

Используя (8.9) и (8.10) из (5.8) находим:

$$\frac{dP^0}{dt} = e^2 \cdot \omega^4 \rho^2 \cdot \left( \frac{14}{15}\gamma^6 - \frac{4}{15}\gamma^4 \right) \quad (8.11)$$

Необходимо прояснить физическую причину различия формул (8.5) и (8.11). Она состоит в том, что (8.5) выражает среднее за период значение потерь энергии частицей, тогда как (8.11) относится к мгновенному значению радиационных потерь. Кажущийся парадокс возникает, из соображения, что при рассматриваемом движении по винтовой линии все точки эквивалентны и излучение частицей энергии не зависит от выбранной точки мировой линии. А раз так, то и среднее значение за период должно совпадать с мгновенным значением потерь.

Однако, последнее утверждение содержит ошибку, так как мгновенное значение потерь энергии вычисляется как поток энергии через сферу, центр которой находит в точек расположения частицы в рассматриваемый момент. Таким образом за период мгновенные потери вычисляются как потоки энергии через *различные* поверхности. Тогда как при вычислении среднего за период значения потерь по формуле (8.5) рассматривается поток энергии через *неподвижную* сферу.

Мы пришли к выводу, что для вычисления среднего по времени значения потерь энергии нельзя пользоваться формулой (5.8), а необходимо вычислить интеграл (4.25). Для вычисления этого интеграла мы применим способ, удобный и в других задачах.

8.2. Центр сферы *S* расположим на оси винтовой линии, по которой движется частица. Тогда

(8.12) $$\gamma^i = \{0;0;0;0\} \quad y^i = z^i$$

Рассмотрим выражение (4.28). Как известно из тензорного анализа свертка градиента функции с некоторым вектором $k^i$ есть производная этой функции по направлению $k^i$. Таким образом, если мы представим

(8.13) $$\zeta_i = w_i + \alpha \cdot w_i + \beta \cdot a_i + \eta \cdot u_i$$

то в (4.28) можно заменить:

(8.14) $$w_m \cdot \frac{\partial}{\partial \zeta_m} \to \frac{\partial}{\partial \alpha}; \quad a_m \cdot \frac{\partial}{\partial \zeta_m} \to \frac{\partial}{\partial \beta}; \quad u_m \cdot \frac{\partial}{\partial \zeta_m} \to \frac{\partial}{\partial \eta}$$

Из (4.28) находим:

(8.15) $$\frac{dP^0}{dt} = \frac{dP^i}{dt} \cdot u_i = -\frac{e^2}{48} \cdot \int_{\sigma_1}^{\sigma_2} d\sigma \cdot \frac{\partial^2}{\partial \eta^2} \left( \frac{\partial^2}{\partial \beta^2} - \gamma^4 \rho^2 \omega^4 \cdot \frac{\partial^2}{\partial \alpha^2} \right) \frac{1}{\left( h^{pq} \zeta_p \zeta_q \right)^{\frac{1}{2}}} \bigg|_{\alpha,\beta,\eta=0}$$

Учитывая (8.8), перейдем к интегрированию по времени лабораторного наблюдателя и вынесем за знак интеграла операторы дифференцирования по нефизическим переменным:

(8.16) $$\frac{dP^0}{dt} = \frac{dP^i}{dt} \cdot u_i = -\frac{e^2}{48 \cdot \gamma} \cdot \frac{\partial^2}{\partial \eta^2} \left( \frac{\partial^2}{\partial \beta^2} - \gamma^4 \rho^2 \omega^4 \cdot \frac{\partial^2}{\partial \alpha^2} \right) \int_{-t_e}^{t_e} dt \cdot \frac{1}{\left( h^{pq} \zeta_p \zeta_q \right)^{\frac{1}{2}}} \bigg|_{\alpha,\beta,\eta=0}$$

Здесь пределы интегрирования по $dt$ должны быть определены из условия (2.29):

(8.17) $$t_e = \frac{\rho}{\sqrt{1-V^2}}$$

7.3. Из (8.1), (8.13) и (4.19) находим:

(8.18) $$h^{pq} \zeta_p \zeta_q = A \cdot t^2 + B \cdot t + C$$

где через A,B,C обозначены следующие выражения:

(8.19) $$A = (1+\alpha)^2 \cdot (1-V^2) \cdot (\gamma^2 \cdot (1-V^2) - 1) + (1-V^2) \cdot \beta^2 \gamma^4 \rho^2 \omega^4 + \eta^2 V^2$$

(8.20) $$B = 2 \cdot ((1+\alpha) \cdot \gamma \cdot (1-V^2) + \eta) \cdot \beta \cdot \gamma^2 \rho^2 \omega^2$$

(8.21) $$C = \rho^2 \cdot ((1+\alpha)^2 + \eta^2 + 2 \cdot (1+\alpha) \cdot \eta \gamma)$$

Не вычисляя дискриминанта (8.18) из (8.19)–(8.21) легко увидеть, что поскольку $\alpha, \beta, \eta$ малы, то $A>0, B\approx 0, C>0$, а следовательно $\Delta = B^2 - 4\cdot A\cdot C < 0$. При этих значениях неопределенный интеграл (8.16) можно представить в виде (см. напр. [4]):

$$(8.22) \quad \int \frac{dt}{\left(At^2 + Bt + C\right)^{\frac{1}{2}}} = \frac{1}{\sqrt{A}} \cdot \ln\left( \frac{2At + B}{\sqrt{-\Delta}} + \sqrt{\frac{(2At+B)^2}{-\Delta} + 1} \right)$$

Для определенного интеграла из (8.22) находим выражение:

$$(8.23) \quad \int_{-t_e}^{t_e} \frac{dt}{\left(At^2 + Bt + C\right)^{\frac{1}{2}}} = \frac{1}{\sqrt{A}} \cdot \ln\left( \frac{2At_e + B + 2\cdot\sqrt{A}\cdot\sqrt{At_e^2 + Bt_e + C}}{-2At_e + B + 2\cdot\sqrt{A}\cdot\sqrt{At_e^2 - Bt_e + C}} \right)$$

Многочлен под знаком радикала в (8.23) есть не что иное, как значение свертки $h^{pq}\zeta_p\zeta_q$ на концах отрезка интегрирования. Как видно из (4.19) и (2.29) многочлен в этих точках представляет собой полный квадрат : $h^{pq}\zeta_p\zeta_q = \left(y^p\zeta_p\right)^2$.

Введем две дополнительные константы:

$$(8.24) \quad D = \beta\cdot\gamma^2\cdot\rho\omega^2; \quad E = (1+\alpha)\cdot\gamma\cdot(1-V^2) + \eta,$$

с помощью которых можно написать:

$$(8.25) \quad A\cdot t_e^2 \pm B\cdot t_e + C = \left(E\cdot t_e \pm D\cdot\rho\right)^2$$

Преобразуем (8.23) с помощью (8.25):

$$(8.26) \quad \int_{-t_e}^{t_e} \frac{dt}{\left(At^2 + Bt + C\right)^{\frac{1}{2}}} = \frac{1}{\sqrt{A}} \cdot \ln\left( \frac{2At_e + 2ED\rho + 2\cdot\sqrt{A}\cdot(Et_e + D\rho)}{-2At_e + 2ED\rho + 2\cdot\sqrt{A}\cdot(Et_e - D\rho)} \right)$$

или, сокращая дробь под знаком логарифма в (8.26):

$$(8.27) \quad \int_{-t_e}^{t_e} \frac{dt}{\left(At^2 + Bt + C\right)^{\frac{1}{2}}} = \frac{1}{\sqrt{A}} \cdot \ln\left( \frac{E + \sqrt{A}}{E - \sqrt{A}} \right)$$

7.4. Остается продифференцировать (8.27) по нефизическим параметрам. Из (8.19) и (8.24) видно, что

$$(8.28) \quad \left.\frac{\partial A}{\partial \beta}\right|_{\beta=0} = 0; \quad \left.\frac{\partial^2 A}{\partial \beta^2}\right|_{\beta=0} = 2\cdot(1-V^2)\cdot\gamma^4\cdot\rho^2\omega^4; \quad \frac{\partial E}{\partial \beta} = 0$$

Учитывая (8.28), дифференцируем (8.27) дважды по $d\beta$:

$$(8.29) \quad \frac{\partial^2}{\partial \beta^2} \int_{-t_e}^{t_e} \frac{dt}{\left(At^2 + Bt + C\right)^{\frac{1}{2}}} = \left( -\frac{1}{2}\frac{1}{A^{3/2}}\cdot\ln\left(\frac{E+\sqrt{A}}{E-\sqrt{A}}\right) + \frac{1}{A}\cdot\frac{E}{E^2-A} \right)\cdot 2\cdot(1-V^2)\cdot\gamma^4\cdot\rho^2\omega^4$$

Для производных величин *A* и *E* по $d\eta$ получим формулы:

(8.30) $$\left.\frac{\partial A}{\partial \eta}\right|_{\eta=0} = 0; \quad \frac{\partial^2 A}{\partial \eta^2} = 2 \cdot V^2; \quad \frac{\partial E}{\partial \eta} = 1$$

Дифференцируя (8.29) дважды по $d\eta$ и учитывая (8.30), находим:

(8.31) $$\frac{\partial^2}{\partial \eta^2}\frac{\partial^2}{\partial \beta^2}\int_{-t_e}^{t_e}\frac{dt}{\left(At^2+Bt+C\right)^{\frac{1}{2}}} = \frac{3V^2}{2}\frac{1}{A^{5/2}} \cdot \ln\left(\frac{E+\sqrt{A}}{E-\sqrt{A}}\right) \cdot 2 \cdot (1-V^2) \cdot \gamma^4 \cdot \rho^2\omega^4$$
$$+ \left(-\frac{3}{A^2}\cdot\frac{V^2 \cdot E}{E^2-A} + \frac{1}{A}\cdot\frac{2EV^2}{(E^2-A)^2} + \frac{8E}{(E^2-A)^3}\right)\cdot 2\cdot(1-V^2)\cdot\gamma^4\cdot\rho^2\omega^4$$

В соответствии с (8.16) необходимо продифференцировать дважды интеграл (8.27) по $d\eta$ и дважды по $d\alpha$. Дифференцируя (8.27) дважды по $d\eta$ и учитывая (8.30), находим:

(8.32) $$\frac{\partial^2}{\partial \eta^2}\int_{-t_e}^{t_e}\frac{dt}{\left(h^{pq}\zeta_p\zeta_q\right)^{\frac{1}{2}}} = -\frac{V^2}{A^{3/2}}\cdot\ln\left(\frac{E+\sqrt{A}}{E-\sqrt{A}}\right) + \frac{2E}{A}\cdot\left(\frac{V^2}{E^2-A}+\frac{2A}{(E^2-A)^2}\right)$$

В (8.32) мы должны положить $\eta=0, \beta=0$ и продифференцировать дважды по $d\alpha$, причём величины *A* и *E* принимают вид:

(8.33) $$A = (1+\alpha)^2 \cdot (1-V^2) \cdot (\gamma^2 \cdot (1-V^2) - 1); \quad E = (1+\alpha) \cdot \gamma \cdot (1-V^2)$$

Из (8.33) видно, что выражение под знаком логарифма в (8.32) не зависит от $\alpha$, а, следовательно, правая часть (8.32) пропорциональна $(1+\alpha)^{-3}$, в связи с чем дифференцирование по $d\alpha$ элементарно и приводит к результату:

(8.34) $$\frac{\partial^2}{\partial \alpha^2}\frac{\partial^2}{\partial \eta^2}\int_{-t_e}^{t_e}\frac{dt}{\left(h^{pq}\zeta_p\zeta_q\right)^{\frac{1}{2}}} = 12\cdot\left(-\frac{V^2}{A^{3/2}}\cdot\ln\left(\frac{E+\sqrt{A}}{E-\sqrt{A}}\right) + \frac{2E}{A}\cdot\left(\frac{V^2}{E^2-A}+\frac{2A}{(E^2-A)^2}\right)\right)$$

Теперь, подставляя (8.31) и (8.34) в (8.16) получим выражение для потери энергии излучающей частицей в единицу времени:

(8.35) $$\frac{dP^0}{dt} = -\frac{e^2\cdot\gamma^3\cdot\rho^2\omega^4}{16}\cdot V^2 \cdot \left(\frac{1-V^2}{A^{5/2}}+\frac{4}{A^{3/2}}\right)\cdot\ln\left(\frac{E+\sqrt{A}}{E-\sqrt{A}}\right) +$$
$$+ \frac{e^2\cdot\gamma^4\cdot\rho^2\omega^4}{24}\cdot\left(\frac{3\cdot V^2\cdot(1-V^2)}{A^2}+\frac{10\cdot V^2}{A}+\frac{16}{1-V^2}\right)$$

Выразим величины *A* и *E* через параметр $\xi$, определённый в (8.6):

(8.36) $$A = (1-V^2)^2\cdot\gamma^2\cdot\xi^2; \quad E = \gamma\cdot(1-V^2)$$

Преобразуя (8.35) с помощью (8.36), найдем:

(8.37)
$$\frac{dP^0}{dt} = \frac{2 \cdot e^2 \cdot \gamma^4 \cdot \rho^2 \omega^4}{3 \cdot (1-V^2)} + \frac{e^2 \cdot \gamma^2 \cdot \rho^2 \omega^4}{16 \cdot (1-V^2)^2} \cdot V^2 \cdot \left( \frac{1}{\xi^5} + \frac{2}{\xi^3} - \frac{3}{\xi} \right) \cdot \ln\left( \frac{1-\xi}{1+\xi} \right) +$$
$$+ \frac{e^2 \cdot \gamma^2 \cdot \rho^2 \omega^4}{16 \cdot (1-V^2)^2} \cdot V^2 \cdot \left( \frac{2}{\xi^4} + \frac{14}{3 \cdot \xi^2} \right)$$

Выражение (8.37) в точности совпадает с результатом (8.5)-(8.7), полученным интегрированием формулы из [1]. Это совпадение служит подтверждением того, что предложенный в данной работе метод вычисления потери энергии движущейся заряженной частицей, приводит к тем же результатам, что и использованный Гинзбургом, Сазоновым и Сыроватским в работах [1].

## Приложение 1.

Вычисление интеграла (5.8).

Из (4.25) видно, что величина $Q^{ikm}$ входит интеграл в составе сверток $Q^{ikm}w_k w_m$ и $Q^{ikm}a_k a_m$. Вычислим их последовательно. Преобразуем первую, используя (5.7):

(9.1) $$Q^{ikm}w_k w_m = \frac{-12 \cdot h^{ik}w_k \cdot h^{mp}u_p w_m + 4 \cdot h^{ip}u_p \cdot U^2}{U^5}$$

Для удобства вычислений введем обозначения

(9.2) $$\xi = \sigma - U$$

(9.3) $$\eta^i = \beta^i + w^i \cdot U$$

Заметим, что

(9.4) $$h^{ik}w_k = \beta^i(\sigma - U) + w^i \sigma U = \eta^i \cdot \xi + w^i \cdot U^2$$

(9.5) $$y^k y_k = \sigma \cdot (\sigma - 2U) = \xi^2 - U^2$$

откуда следует, что

(9.6) $$Q^{ikm}w_k w_m = -12 \cdot \frac{(\eta^i \cdot \xi + w^i \cdot U^2) \cdot (\eta^p \cdot \xi + w^p \cdot U^2) \cdot u_p}{U^5}$$
$$+ 4 \cdot \frac{\eta^i \cdot (\eta^p u_p + w^p u_p \cdot \xi) + w^i \cdot (\eta^p u_p \cdot \xi + w^p u_p \cdot \xi^2) - u^i \cdot (\xi^2 - U^2)}{U^3}$$

При вычислении интеграла нечетные степени $\xi$ дают нулевой вклад:

(9.7) $$\int_{-U}^{U} Q^{ikm}w_k w_m d\xi = -\frac{12}{U^5} \cdot \int_{-U}^{U} (\eta^i \eta^p u_p \cdot \xi^2 + w^i w^p u_p \cdot U^4) d\xi$$
$$+ \frac{4}{U^3} \cdot \int_{-U}^{U} ((\eta^i \eta^p u_p + w^i w^p u_p \cdot \xi^2) - u^i \cdot (\xi^2 - U^2)) d\xi$$

Вычисление интегралов (9.7) приводит к результату

(9.8) $$\int_{-U}^{U} Q^{ikm}w_k w_m d\xi = -\frac{64}{3} w^i \cdot w^p u_p + \frac{16}{3} u^i$$

Для вычисления свертки $Q^{ikm}a_k a_m$ введем обозначения:

(9.9) $$A = \beta^m a_m$$

Ясно, что

(9.10) $$y^m a_m = A = const$$

Из (9.4) видно, что

(9.11) $$h^{ik}a_i w_k = A \cdot (\sigma - U) = A \cdot \xi$$

Учитывая (9.9) – (9.11) перепишем свертку в виде:

(9.12) $$Q^{ikm}a_k a_m = -\frac{2h^{mp}h^{ik} + h^{ip}h^{km}}{U^5}u_p a_k a_m + 5 \cdot \frac{2(h^{pk}h^{iq} + h^{ik}h^{pq})a_k w_q \cdot A \cdot \xi}{U^7}u_p + $$
$$+5 \cdot \frac{h^{pq}w_q \cdot h^{km}h^{ir} \cdot a_m \cdot a_k \cdot w_r + h^{ip} \cdot A^2 \cdot \xi^2}{U^7} \cdot u_p - 35 \cdot \frac{h^{pq}w_q \cdot h^{il}w_l}{U^9} \cdot u_p \cdot A^2 \xi^2$$

Свертка (9.12) – вектор, составленный из векторов $\beta^i, w^i, a^i, u^i$. Представим его в виде:

(9.13) $$Q^{ikm}a_k a_m = h^{iq} \cdot (w_q \cdot XW + a_q \cdot XA + u_q \cdot XU)$$

где

(9.14) $$XW = 5 \cdot \frac{2h^{pk}a_k \cdot A \cdot \xi}{U^7}u_p + 5 \cdot \frac{h^{pq}w_q \cdot h^{km} \cdot a_m \cdot a_k}{U^7} \cdot u_p - 35 \cdot \frac{h^{pq}w_q}{U^9} \cdot u_p \cdot A^2\xi^2$$

(9.15) $$XA = -\frac{2h^{mp}}{U^5}u_p a_m + 5 \cdot \frac{2h^{pq}w_q \cdot A \cdot \xi}{U^7}u_p$$

(9.16) $$XU = -\frac{h^{km}}{U^5}a_k a_m + 5 \cdot \frac{A^2 \cdot \xi^2}{U^7}$$

Разобьем $XW$ на три слагаемых

(9.17) $$XW = XW_1 + XW_2 + XW_3$$

и проинтегрируем по отдельности

(9.18) $$IW_1^i = \int_{-U}^{U} h^{ik}w_k \cdot 5 \cdot \frac{2h^{pk}a_k \cdot A \cdot \xi}{U^7}u_p \cdot d\xi$$

(9.19) $$IW_2^i = \int_{-U}^{U} h^{ik}w_k \cdot 5 \cdot \frac{h^{pq}w_q \cdot h^{km} \cdot a_m \cdot a_k}{U^7} \cdot u_p \cdot d\xi$$

(9.20) $$IW_3^i = -\int_{-U}^{U} h^{ik}w_k \cdot 35 \cdot \frac{h^{pq}w_q}{U^9} \cdot u_p \cdot A^2\xi^2 \cdot d\xi$$

Вычисляем интегралы (9.18)-(9.20) и складываем, обозначив сумму через $IW^i$:

(9.21) $$IW^i = \frac{8A}{3U^2} \cdot \eta^i \cdot a^p u_p + \eta^i \eta^p u_p \cdot \left(\frac{4}{3} \cdot \frac{1}{U^2} \cdot a^n a_n - \frac{4A^2}{U^4}\right) + w^i w^p u_p \cdot \left(\frac{20}{3} \cdot a^n a_n - \frac{20A^2}{3U^2}\right)$$

Затем вычисляем интегралы

(9.22) $$IA^i = \int_{-U}^{U} h^{ik}a_k \cdot \left(-\frac{2h^{mp}}{U^5}u_p a_m + 5 \cdot \frac{2h^{pq}w_q \cdot A \cdot \xi}{U^7}u_p\right) \cdot d\xi$$

(9.23)
$$IU^i = \int_{-U}^{U} h^{ik} u_k \cdot \left( -\frac{h^{km}}{U^5} a_k a_m + 5 \cdot \frac{A^2 \cdot \xi^2}{U^7} \right) \cdot d\xi$$

Вычисление приводит к результатам:

(9.24)
$$IA^i = +\frac{8}{3}\frac{A^2}{U^4} \cdot \eta^i \eta^p u_p + \frac{16 A^2}{3U^2} w^p w^i u_p - \frac{8 A u_p}{3U^2}\left(\eta^i a^p\right) - \frac{32}{15} a^i a^p u_p$$

(9.25)
$$IU^i = \eta^i \eta^p u_p \left( \frac{4}{3} \cdot \frac{A^2}{U^4} - \frac{4}{3} \frac{a^n a_n}{U^2} \right) + w^i w^p u_p \left( \frac{4}{3} \cdot \frac{A^2}{U^2} - \frac{4}{15} a^n a_n \right) + u^i \cdot \frac{16}{15} a^n a_n ( )$$

Суммируя (9.21),(9.24) и (9.25) находим:

(9.26)
$$\int_{-U}^{U} Q^{ikm} a_k a_m d\xi = + w^i w^p u_p \cdot \frac{32}{5} \cdot a^n a_n - \frac{32}{15} a^i a^p u_p - u^i \cdot \frac{16}{15} a^n a_n$$

Подставив (9.8) и (9.26), приведем выражение (4.26) окончательному виду:

(9.27)
$$\frac{dP^i}{dt} = e^2 \cdot \left( -\frac{14}{15} w^i w^p u_p \cdot a^n a_n - \frac{2}{15} \cdot a^i \cdot a^p u_p + \frac{4}{15} \cdot u^i \cdot a^n a_n \right)$$

## Приложение 2.

Вычисление потерь энергии частицы при движении в магнитном поле.

Формула Гинзбурга, Сазонова, Сыроватского в наших обозначениях принимает вид:

$$\vec{E}_n = \frac{2e}{r} \cdot \frac{n\rho \omega_H^2}{(1-Vk_z)^2} \left\{ \vec{l}_1 \cdot J'_n(z_n) - i\vec{l}_2 \cdot \frac{k_z - V}{\rho \omega_H \cdot k_\perp} \cdot J_n(z_n) \right\} \tag{10.1}$$

где

$$z_n = n \cdot \frac{\rho \omega_H k_\perp}{1 - V k_z} \tag{10.2}$$

$k_z, k_\perp$ — компоненты единичного вектора $\vec{k}$, направленного на точку наблюдения;

$\vec{l}_1, \vec{l}_2$ — единичные векторы, ортогональные $\vec{k}$; $J_n(z)$ — функция Бесселя.

Средняя за период плотность потока вектора Пойнтинга:

$$S_n = \frac{1}{8\pi} \cdot \vec{E}_n \cdot \vec{E}_n^* \tag{10.3}$$

Суммируя по частотам, находим среднюю плотность потока энергии в направлении $\vec{k}$:

$$S = \sum_{n=1}^{\infty} \frac{e^2}{2\pi r^2} \cdot \frac{\rho^2 \omega_H^4}{(1-Vk_z)^4} \left\{ n^2 \left(J'_n(n\xi)\right)^2 + \left(\frac{k_z - V}{\rho\omega_H \cdot k_\perp}\right)^2 \cdot n^2 \left(J_n(n\xi)\right)^2 \right\} \qquad (10.4)$$

Воспользуемся известными формулами для сумм рядов [5, с. 102][1]:

$$\sum_{n=1}^{\infty} n^2 J_n^2(nz) = \frac{z^2(4+z^2)}{16 \cdot (1-z^2)^{\frac{7}{2}}} \qquad (10.5)$$

$$\sum_{k=1}^{\infty} n^2 J_n^{'2}(nz) = \frac{4+3z^2}{16 \cdot (1-z^2)^{\frac{5}{2}}} \qquad (10.6)$$

Итак, суммарный поток на всех частотах:

$$S = \frac{e^2}{2\pi r^2} \cdot \frac{\rho^2 \omega_H^4}{(1-Vk_z)^4} \left\{ \frac{(4+3\zeta^2)}{16 \cdot (1-\zeta^2)^{\frac{5}{2}}} + \left(\frac{k_z - V}{\rho\omega_H \cdot k_\perp}\right)^2 \cdot \frac{\zeta^2 \cdot (4+\zeta^2)}{16 \cdot (1-\zeta^2)^{\frac{7}{2}}} \right\} \qquad (10.7)$$

где

$$\zeta = \frac{\rho\omega_H k_\perp}{1-Vk_z} \qquad (10.8)$$

Для вычисления потери энергии частицей проинтегрируем (10.7) по поверхности сферы, для чего введем сферические координаты с полярной осью, направленной вдоль магнитного поля. Тогда

$$k_z = \cos\vartheta; \quad k_\perp = \sin\vartheta \qquad (10.9)$$

Для полной энергии, излучаемой в единицу времени:

$$\frac{dP^0}{dt} = \int_0^{2\pi} d\varphi \int_0^\pi r^2 S \cdot \sin\theta' d\theta' \qquad (10.10)$$

Выполняя в (10.10) тривиальное интегрирование по $d\varphi$ и заменяя переменную интегрирования $u = \cos\vartheta$, из (10.7)-(10.8) получим:

$$\frac{dP^0}{dt} = \frac{e^2 \rho^2 \omega_H^4}{16} \cdot \int_{-1}^{1} \frac{du}{(1-uV)} \left\{ -\frac{3}{(P(u))^{\frac{3}{2}}} + \frac{7 \cdot (1-uV)^2 - (u-V)^2}{(P(u))^{\frac{5}{2}}} + \frac{5 \cdot (u-V)^2 \cdot (1-uV)^2}{(P(u))^{\frac{7}{2}}} \right\} \qquad (10.11)$$

где

$$P(u) = (V^2 + \rho^2 \omega_H^2) \cdot u^2 - 2uV + (1 - \rho^2 \omega_H^2) \qquad (10.12)$$

Интеграл в (10.11) распадается на два, которые мы вычислим независимо:

$$I_1 = \int_{-1}^{1} du \left\{ \frac{7 \cdot (1-uV)^2}{(P(u))^{\frac{5}{2}}} + \frac{5 \cdot (u-V)^2 \cdot (1-uV)}{(P(u))^{\frac{7}{2}}} \right\} \qquad (10.13)$$

---

[1] Попутно отметим, что в справочнике [6] формула (10.5) на стр. 689 дана с опечаткой.

$$I_2 = \int_{-1}^{1} \frac{du}{(1-uV)} \left\{ \frac{-3P(u)-(u-V)^2}{(P(u))^{\frac{5}{2}}} \right\} \qquad (10.14)$$

Начнем вычисление с интеграла $I_1$.

Числитель второго слагаемого в фигурных скобках (10.13) преобразуем к виду:

$$(u-V)^2 \cdot (1-uV) = -\frac{1}{A} \cdot \left( Vu^2 - \left(2V^2 + \frac{W^2}{A}\right) \cdot u + A \cdot V \right) \cdot (A \cdot u - V) + \frac{W^2 \cdot V}{A^2 \cdot \gamma^4} \cdot u \qquad (10.15)$$

где через $W$ обозначена перпендикулярная полю составляющая скорости частицы $W = \rho \cdot \omega_H$, через $A$ – старший коэффициент многочлена $P(u)$:

$$A = V^2 + W^2 \qquad (10.16)$$

а $\gamma$ – релятивистский фактор:

$$\gamma = \frac{1}{\sqrt{1-V^2-W^2}} \qquad (10.17)$$

Учитывая, что $A \cdot u - V = \frac{1}{2} P'(u)$, второе слагаемое в (10.13) может быть проинтегрировано по частям:

$$I_1 = \int_{-1}^{1} du \left\{ \frac{7 \cdot (1-uV)}{(P(u))^{\frac{5}{2}}} + 5 \cdot \frac{W^2 \cdot V}{A^2 \cdot \gamma^4} \cdot \frac{u}{(P(u))^{\frac{7}{2}}} \right\} - \frac{1}{A} \cdot \int_{-1}^{1} \frac{du}{(P(u))^{\frac{5}{2}}} \cdot \left( 2Vu - \left(2V^2 + \frac{W^2}{A}\right) \right)$$
$$+ \frac{1}{A} \cdot \left( P^{-5/2} \right) \cdot \left( Vu^2 - \left(2V^2 + \frac{W^2}{A}\right) \cdot u + A \cdot V \right) \Big|_{-1}^{1} \qquad (10.18)$$

Прямым дифференцированием может быть проверена следующая формула:

$$\int du \frac{u}{(P(u))^{\frac{n}{2}}} = -\frac{1-W^2-Vu}{W^2} \cdot \frac{\gamma^2}{(n-2)} \cdot \frac{1}{(P(u))^{\frac{n}{2}-1}} + \frac{V \cdot \gamma^2}{W^2} \cdot \frac{n-3}{n-2} \int \frac{du}{(P(u))^{\frac{n}{2}-1}} \qquad (10.19)$$

Применив (10.19) ко второму слагаемому в фигурных скобках (10.18), получим:

$$I_1 = \int_{-1}^{1} du \left\{ \frac{7 \cdot (1-uV)}{(P(u))^{\frac{5}{2}}} + 4 \cdot \frac{V^2}{A^2 \cdot \gamma^2} \cdot \frac{1}{(P(u))^{\frac{5}{2}}} \right\} - \frac{1}{A} \cdot \int_{-1}^{1} \frac{du}{(P(u))^{\frac{5}{2}}} \cdot \left( 2Vu - \left(2V^2 + \frac{W^2}{A}\right) \right)$$
$$- \frac{V}{A^2 \cdot \gamma^2} \cdot (1-W^2-Vu) \cdot \frac{1}{(P(u))^{\frac{5}{2}}} + \frac{1}{A} \cdot \left( Vu^2 - \left(2V^2 + \frac{W^2}{A}\right) \cdot u + A \cdot V \right) \cdot \frac{1}{(P(u))^{\frac{5}{2}}} \Big|_{-1}^{1} \qquad (10.20)$$

Приведя в (10.20) подобные члены, придем к выражению:

$$I_1 = M_1 \cdot \int_{-1}^{1} \frac{du}{(P(u))^{\frac{5}{2}}} + N_1 \cdot \int_{-1}^{1} \frac{u \cdot du}{(P(u))^{\frac{5}{2}}} + Q_1 \qquad (10.21)$$

где введены обозначения

$$M_1 = 7 + \frac{4V^2}{A^2 \cdot \gamma^2} + \frac{2V^2}{A} + \frac{W^2}{A^2} \qquad (10.22)$$

$$N_1 = -7V - \frac{2V}{A} \qquad (10.23)$$

$$Q_1 = \left[ -\frac{V}{A^2 \cdot \gamma^2} \cdot \left(1 - W^2 - Vu\right) + \frac{1}{A} \cdot \left(Vu^2 - \left(2V^2 + \frac{W^2}{A}\right) \cdot u + A \cdot V\right) \right] \cdot \frac{1}{(P(u))^{\frac{5}{2}}} \Bigg|_{-1}^{1} \qquad (10.24)$$

После упрощающих преобразований величина $Q_1$ принимает вид:

$$Q_1 = -\frac{2}{A} \cdot \frac{1+V^2}{\left(1-V^2\right)^2} - \frac{2V^2}{A^2} \cdot \frac{3+V^2}{\left(1-V^2\right)^3} \qquad (10.25)$$

Следующая формула проверяется прямым дифференцированием:

$$\int \frac{du}{(P(u))^{\frac{n}{2}}} = \frac{A \cdot u - V}{W^2} \cdot \frac{\gamma^2}{(n-2)} \cdot \frac{1}{(P(u))^{\frac{n}{2}-1}} + \frac{A \cdot \gamma^2}{W^2} \cdot \frac{n-3}{n-2} \int \frac{du}{(P(u))^{\frac{n}{2}-1}} \qquad (10.26)$$

С помощью (10.19) и (10.26) выражение (10.21) преобразуем к виду:

$$I_1 = M_1 \cdot \frac{A \cdot \gamma^2}{W^2} \cdot \frac{2}{3} \cdot \int_{-1}^{1} \frac{du}{(P(u))^{\frac{3}{2}}} + N_1 \cdot \frac{V \cdot \gamma^2}{W^2} \cdot \frac{2}{3} \int_{-1}^{1} \frac{du}{(P(u))^{\frac{3}{2}}} + Q_1 + Q_2 \qquad (10.27)$$

где

$$Q_2 = \left( M_1 \cdot \frac{A \cdot u - V}{W^2} \cdot \frac{\gamma^2}{3} - N_1 \cdot \frac{1 - W^2 - Vu}{W^2} \cdot \frac{\gamma^2}{3} \right) \cdot \frac{1}{(P(u))^{\frac{3}{2}}} \Bigg|_{-1}^{1} \qquad (10.28)$$

С помощью (10.26) преобразуем (10.27):

$$I_1 = \frac{2 \cdot \gamma^2}{3 \cdot W^2} \cdot \left( M_1 \cdot A + N_1 \cdot V \right) \cdot \left( \frac{A \cdot u - V}{W^2} \cdot \frac{\gamma^2}{(P(u))^{\frac{1}{2}}} \right) \Bigg|_{-1}^{1} + Q_1 + Q_2 \qquad (10.29)$$

Упрощая (10.28) и складывая с (10.25), после упрощающих преобразований получим:

$$Q_1 + Q_2 = \frac{20 \cdot \gamma^2}{3} \cdot \frac{1+V^2}{\left(1-V^2\right)^2} - \frac{4\gamma^2}{3 \cdot W^2 \cdot A} \cdot \frac{2 \cdot V^2 + W^2}{\left(1-V^2\right)} \qquad (10.30)$$

Подставив выражения (10.22), (10.23), (10.30) в (10.29), после упрощения получим

$$I_1 = \frac{20 \cdot \gamma^2}{3} \cdot \frac{1+V^2}{\left(1-V^2\right)^2} + \frac{32 \cdot \gamma^4}{3} \cdot \frac{1}{1-V^2} \qquad (10.31)$$

Обратимся теперь к вычислению интеграла (10.14). С помощью замены переменной интегрирования $u = \dfrac{1}{V} + x$ приведем интеграл к виду

$$I_2 = \frac{1}{V} \int_{x_1}^{x_2} \frac{dx}{x} \left\{ \frac{M_2 x^2 + N_2 x + K_2}{\left(P_1(x)\right)^{\frac{5}{2}}} \right\} \tag{10.32}$$

где $x_1 = -1 - \dfrac{1}{V}$; $x_2 = 1 - \dfrac{1}{V}$.

Через $P_1(x)$ обозначен многочлен

$$P_1(x) = A \cdot x^2 + 2x \cdot \frac{W^2}{V} + \frac{W^2}{V^2} - W^2 \tag{10.33}$$

Коэффициенты многочлена в числителе (10.32) имеют следующие значения:

$$M_2 = 3 \cdot A + 1 \tag{10.34}$$

$$N_2 = \frac{2}{V} \cdot \left(3 \cdot W^2 + 1 - V^2\right) \tag{10.35}$$

$$K_2 = \frac{1 - V^2}{V^2} \cdot \left(3W^2 + 1 - V^2\right) \tag{10.36}$$

Таким образом, интеграл (10.32) распадается на три:

$$I_2 = \frac{M_2}{V} \cdot \int_{x_1}^{x_2} \frac{x\,dx}{\left(P_1(x)\right)^{\frac{5}{2}}} + \frac{N_2}{V} \cdot \int_{x_1}^{x_2} \frac{dx}{\left(P_1(x)\right)^{\frac{5}{2}}} + \frac{K_2}{V} \cdot \int_{x_1}^{x_2} \frac{dx}{x \cdot \left(P_1(x)\right)^{\frac{5}{2}}} \tag{10.37}$$

Следующая формула проверяется дифференцированием:

$$\int \frac{dx}{x \cdot \left(P_1\right)^{\frac{n}{2}}} = \frac{1}{1 - V^2} \cdot \left( -V \cdot \int \frac{dx}{\left(P_1\right)^{\frac{n}{2}}} + \frac{V^2}{W^2} \cdot \left( \frac{1}{(n-2) \cdot \left(P_1\right)^{\frac{n}{2}-1}} + \int \frac{dx}{x \cdot \left(P_1\right)^{\frac{n}{2}-1}} \right) \right) \tag{10.38}$$

Применяя (10.38) к последнему члену (10.37), найдем:

$$I_2 = \frac{M_2}{V} \cdot \int_{x_1}^{x_2} \frac{x\,dx}{\left(P_1(x)\right)^{\frac{5}{2}}} + \left( \frac{N_2}{V} - \frac{K_2}{1 - V^2} \right) \cdot \int_{x_1}^{x_2} \frac{dx}{\left(P_1(x)\right)^{\frac{5}{2}}}$$
$$+ \frac{K_2 \cdot V}{W^2 \cdot (1 - V^2)} \cdot \int_{x_1}^{x_2} \frac{dx}{x \cdot \left(P_1(x)\right)^{\frac{3}{2}}} + \frac{K_2 \cdot V}{W^2 \cdot (1 - V^2)} \cdot \frac{1}{3 \cdot \left(P_1(x)\right)^{\frac{3}{2}}} \bigg|_{x_1}^{x_2} \tag{10.39}$$

Еще две формулы для первообразных:

$$\int \frac{x \cdot dx}{\left(P_1(x)\right)^{\frac{n}{2}}} = -\left( \frac{\gamma^2}{V} x + \frac{(1 - V^2)\gamma^2}{V^2} \right) \cdot \frac{1}{(n - 2)} \cdot \frac{1}{\left(P_1(x)\right)^{\frac{n}{2}-1}} - \frac{\gamma^2}{V} \frac{n - 3}{n - 2} \int \frac{dx}{\left(P_1(x)\right)^{\frac{n}{2}-1}} \tag{10.40}$$

$$\int \frac{dx}{(P_1(x))^{\frac{n}{2}}} = \left(\frac{A\gamma^2}{W^2}x + \frac{\gamma^2}{V}\right) \cdot \frac{1}{(n-2)} \cdot \frac{1}{(P_1(x))^{\frac{n}{2}-1}} + \frac{A\gamma^2}{W^2} \frac{n-3}{n-2} \int \frac{dx}{(P_1(x))^{\frac{n}{2}-1}} \qquad (10.41)$$

С помощью (10.38), (10.40) и (10.41) формула (10.39) приводится к виду:

$$I_2 = -\frac{\gamma^2 \cdot M_2}{V^2} \cdot \frac{2}{3} \cdot \int_{x_1}^{x_2} \frac{dx}{(P_1(x))^{\frac{3}{2}}} + \left(\frac{N_2}{V} - \frac{K_2}{1-V^2}\right) \cdot \frac{\gamma^2 A}{W^2} \cdot \frac{2}{3} \cdot \int_{x_1}^{x_2} \frac{dx}{(P_1(x))^{\frac{3}{2}}}$$
$$+ \frac{K_2 \cdot V}{W^2 \cdot (1-V^2)^2} \cdot \left(-V \cdot \int \frac{dx}{(P_1(x))^{\frac{3}{2}}} + \frac{V^2}{W^2} \cdot \int \frac{dx}{x \cdot (P_1(x))^{\frac{1}{2}}}\right) + Q_3 \qquad (10.42)$$

где

$$Q_3 = +\frac{K_2 \cdot V}{W^2 \cdot (1-V^2)} \cdot \frac{1}{3 \cdot (P_1(x))^{\frac{3}{2}}} - \frac{\gamma^2 M_2}{3V} \cdot \left(\frac{x}{V} + \frac{(1-V^2)}{V^2}\right) \cdot \frac{1}{(P_1(x))^{\frac{3}{2}}}\bigg|_{x_1}^{x_2}$$
$$+ \left(\frac{N_2}{V} - \frac{K_2}{1-V^2}\right) \cdot \left(\frac{A\gamma^2}{W^2}x + \frac{\gamma^2}{V}\right) \cdot \frac{1}{3} \cdot \frac{1}{(P_1(x))^{\frac{3}{2}}} + \frac{K_2 \cdot V^3}{W^4 \cdot (1-V^2)^2} \cdot \frac{1}{(P_1(x))^{\frac{1}{2}}}\bigg|_{x_1}^{x_2} \qquad (10.43)$$

Или

$$I_2 = M_3 \cdot \int_{x_1}^{x_2} \frac{dx}{(P_1(x))^{\frac{3}{2}}} + \frac{K_2 \cdot V^3}{W^4 \cdot (1-V^2)^2} \cdot \int_{x_1}^{x_2} \frac{dx}{x \cdot (P_1(x))^{\frac{1}{2}}} + Q_3 \qquad (10.44)$$

где

$$M_3 = -\frac{2 \cdot \gamma^2 \cdot M_2}{3 \cdot V^2} + \frac{2 \cdot \gamma^2 A}{3 \cdot W^2} \cdot \left(\frac{N_2}{V} - \frac{K_2}{1-V^2}\right) - \frac{K_2 \cdot V^2}{W^2 \cdot (1-V^2)^2} \qquad (10.45)$$

Интеграл в первом слагаемом (10.44) вычисляется с помощью (10.41) при $n=3$, а первообразная второго имеет вид:

$$\int \frac{dx}{x \cdot (P_1(x))^{\frac{1}{2}}} = -\frac{V}{W\sqrt{1-V^2}} \cdot \ln\left|\frac{\frac{W}{V} \cdot (1-V^2) + Wx + \sqrt{(1-V^2) \cdot P_1(x)}}{x}\right| \qquad (10.46)$$

Таким образом, интеграл $I_2$ можно представить в виде суммы:

$$I_2 = Q_3 + Q_4 + Q_5 \qquad (10.47)$$

Где

$$Q_4 = M_3 \cdot \left(\frac{A\gamma^2}{W^2}x + \frac{\gamma^2}{V}\right) \cdot \frac{1}{(P_1(x))^{\frac{1}{2}}}\bigg|_{x_1}^{x_2} \qquad (10.48)$$

$$Q_5 = -\frac{K_2 \cdot V^4}{W^5 \cdot \left(1-V^2\right)^{\frac{5}{2}}} \cdot \ln\left|\frac{\frac{W}{V}\cdot\left(1-V^2\right)+Wx+\sqrt{\left(1-V^2\right)\cdot P_1(x)}}{x}\right|\Bigg|_{x_1}^{x_2} \qquad (10.49)$$

Осталось привести подобные члены в формулах (10.43), (10.48), (10.49). Преобразование (10.43), (10.48) и (10.49) приводит к выражениям:

$$Q_3 = \frac{2\cdot V^2}{W^4 \cdot \left(1-V^2\right)} + \frac{1}{3\cdot W^2}\cdot\frac{2+20V^2}{\left(1-V^2\right)^2} \qquad (10.50)$$

$$Q_4 = -\frac{2\gamma^2}{3\cdot W^2 \cdot \left(1-V^2\right)} - \frac{6\gamma^2}{\left(1-V^2\right)^2} \qquad (10.51)$$

$$Q_5 = \frac{\left(3\cdot W^2 + 1 - V^2\right)\cdot V^2}{W^5 \cdot \left(1-V^2\right)^{\frac{3}{2}}}\cdot\ln\left|\frac{1-\xi}{1+\xi}\right| \qquad (10.52)$$

где $\xi = \dfrac{W}{\sqrt{1-V^2}}$.

Из (10.31), (10.50), (10.51), (10.52) получаем после упрощений:

$$I_1 + I_2 = \frac{32\cdot\gamma^4}{3}\cdot\frac{1}{1-V^2} + \frac{V^2\cdot\gamma^2}{\left(1-V^2\right)^2}\cdot\left(\frac{2}{\xi^4}+\frac{14}{3\xi^2}-\left(\frac{3}{\xi}-\frac{2}{\xi^3}-\frac{1}{\xi^5}\right)\cdot\ln\left|\frac{1-\xi}{1+\xi}\right|\right) \qquad (10.53)$$

Из (10.11) находим выражение для радиационных потерь частицы:

$$\frac{dP^0}{dt} = \frac{e^2\rho^2\omega_H^4}{1-V^2}\cdot\left(\frac{2}{3}\gamma^4 + \frac{V^2\gamma^2}{1-V^2}\cdot\Psi(\xi)\right) \qquad (10.54)$$

где $\Psi(\xi)$ – следующая функция одной переменной, определенная на интервале $(0,1)$:

$$\Psi(\xi) = \frac{1}{16}\cdot\left(\frac{2}{\xi^4}+\frac{14}{3\xi^2}-\left(\frac{3}{\xi}-\frac{2}{\xi^3}-\frac{1}{\xi^5}\right)\cdot\ln\left|\frac{1-\xi}{1+\xi}\right|\right) \qquad (10.55)$$

Функция $\Psi(\xi)$ монотонно возрастает от $\dfrac{4}{15}$ (при $\xi = 0$) к $\dfrac{5}{12}$ (при $\xi = 1$).